\documentclass[journal,article,submit,moreauthors]{mdpi} 
\UseRawInputEncoding
\firstpage{1} 
\hreflink{https://doi.org/} 

\usepackage{wrapfig}
\usepackage[pscoord]{eso-pic}
\usepackage[fulladjust]{marginnote}
\usepackage{tikz}
\usepackage[many]{tcolorbox}
\tcbuselibrary{skins, breakable, theorems}

\usepackage{varwidth}
\usepackage{extarrows}
\usepackage{tikz}
\usepackage{tabu}
\usetikzlibrary{arrows}
\usepackage{tcolorbox}
\tcbuselibrary{skins, breakable, theorems}
\usepackage{fancybox}
\usepackage{amsthm}

\reversemarginpar

\definecolor{Gray}{gray}{.25}
\definecolor{lightgray1}{gray}{0.95}
\definecolor{lightgray2}{gray}{0.6}
\definecolor{lightgray3}{gray}{0.8}

\Title{\boldmath Duality family of scalar field} 

\Author{Wen-Du Li $^{1,3,\dagger}$ and Wu-Sheng Dai $^{2,\ddagger}$}

\address{%
$^{1}$ \quad College of Physics and Materials Science, Tianjin Normal University, Tianjin 300387, PR China\\
$^{2}$ \quad Department of Physics, Tianjin University, Tianjin 300350, P.R. China\\
$^{3}$ \quad Theoretical Physics Division, Chern Institute of Mathematics, Nankai University, Tianjin, 300071, P. R. China
}

\firstnote{liwendu@tjnu.edu.cn.} 
\secondnote{daiwusheng@tju.edu.cn.}

\abstract{We show that there exists a duality family of self-interacting massive scalar
fields. The scalar field in a duality family are related by a duality
transformation. Such a duality of scalar fields is a field version of the
Newton-Hooke duality in classical mechanics. The duality transformation
preserves the type of the field equation: transforming a Klein-Gordon type
equation to another Klein-Gordon type equation with a different
self-interacting potential. Once a field in a duality family is solved, all
other family members are solved by the transformation. That is, a series of
exactly solvable models can be constructed from one exactly solvable model.
The dual field of the power-interaction field, the sine-Gordon field, etc.,
are considered. Moreover, as a comparison, we show an analogue of the duality
in classical and quantum mechanics.}

\keyword{Duality family; Exact solution of field equation.} 

\usepackage{titletoc}   
\titlecontents{section}[0.6em]{\vspace{.25\baselineskip}}
             {\thecontentslabel\quad}{}
             {\hspace{.5em}\titlerule*[10pt]{$\cdot$}\contentspage}
\titlecontents{subsection}[2.1em]{\vspace{.25\baselineskip}}
             {\thecontentslabel\quad}{}
             {\hspace{.5em}\titlerule*[10pt]{$\cdot$}\contentspage}

\begin{document}
\nolinenumbers

\tableofcontents



\section{Introduction}

The scalar field with the self-interaction potential $V\left(  \phi\right)  $
determined by the Klein-Gordon type equation%

\begin{equation}
\square\phi+m^{2}\phi+\frac{\partial V\left(  \phi\right)  }{\partial\phi}=0
\label{fieldeqVphi}%
\end{equation}
is an important field-theory model, such as the $\phi^{n}$-field and the
sine-Gordon field. The field equation (\ref{fieldeqVphi}) is usually a
nonlinear equation. The solution of such field equations, e.g., the soliton
solution, is important in studying the nonperturbation aspect of fields.

In this paper, we show that there exists a duality family of self-interaction
scalar fields. The family members are connected by a duality transform. The
duality transformation transforms a Klein-Gordon type equation to another
Klein-Gordon type equation, while, for comparison, a Klein-Gordon type
equation after an arbitrary transformation is usually no longer a Klein-Gordon
type equation. Two dual fields are related by a duality transformation. Once a
field equation is solved, the solution of its dual field can be obtained by
the dual transformation immediately. A field has not only one dual fields. All
fields who are dual to each other form a duality family. Every field belongs
to a certain duality family. A duality family consists of an infinite number
of family members. Different family members may have different masses and
different self-interaction potentials. So long as one field in the duality
family is solved, all other fields in the duality family are solved by the
duality transformation. For example,\ starting with a solution of the
sine-Gordon equation, we can solve all the fields in the sine-Gordon-field
duality family. The duality of the power-interaction field, the sine-Gordon
field, the sinh-Gordon field, etc., are considered as examples.

In order to help understand the duality of scalar fields, we can refer to its
analogue in classical mechanics. In classical mechanics, Newton discovered a
duality in his \textit{Principia\ }(Corollary III of Proposition VII)\textit{
}\cite{chandrasekhar1995newton}. Newton considered such a problem: for a power
law of centripetal attraction, does there exist a dual law for which a body
with the same constant of area describes the same orbit
\cite{chandrasekhar1995newton}. Newton's result is today known as\ the
Newton-Hooke duality, for it is a duality between the Newtonian gravitational
potential ($1/r$-potential) and the Hookian harmonic-oscillator potential
($r^{2}$-potential). Chandrasekhar reexpressed the Newton-Hooke duality in a
more modern language \cite{chandrasekhar1995newton}. E. Kasner and V. I.
Arnol'd independently generalized the Newton-Hooke duality to general power
potentials: two power potentials $U\left(  r\right)  =\xi r^{a}$ and $V\left(
r\right)  =\eta r^{A}$ are dual, if $\frac{a+2}{2}=\frac{2}{A+2}$
\cite{arnold2013mathematical,arnold1990huygens,needham1998visual}. T. Needham
intuitively explains the Newton-Hooke duality and its generalization, the
Kasner-Arnol'd theorem, in Refs. \cite{needham1998visual,needham1993newton}.
R. W. Hall and K. Josic reviewed the power-potential duality in Ref.
\cite{hall2000planetary}. In appendix \ref{travelingwavesolution}, we
generalize the classical mechanical general power-potential duality to
arbitrary potentials in classical mechanics and in quantum mechanics in
arbitrary dimensions.

Various dualities reveal underlying connections among different physical
problems. The gauge/gravity duality\ is a profound relation
\cite{maldacena1997large,witten1998anti,witten1998anti2} and has been applied
in many problems
\cite{aharony2000large,d2004supersymmetric,perez2016boundary,garcia2017digital,hashimoto2018deep,de2018chaotic,bigazzi2018neutron,zangeneh2016thermodynamics}%
. The fluid/gravity duality is a duality between spacetime manifolds and
fluids
\cite{bredberg2012navier,hubeny2011fluid,compere2011holographic,hao2015flat,quiros2000dual,bhattacharyya2008nonlinear,bhattacharyya2009forced,bhattacharyya2008conformal,ashok2014forced,bhattacharyya2009incompressible,compere2012relativistic,pinzani2015towards,wu2013fluid}%
. The gravoelectric duality is useful to seek the solution of the Einstein
equation
\cite{dadhich2002most,nouri1999spacetime,dadhich2000electromagnetic,dadhich1998duality,dadhich2000gravoelectric}%
. The strong--weak duality bridges a strongly coupled theory to an equivalent
weak coupling theory: the duality between fermions and strongly-interacting
bosonic Chern-Simons-matter theories \cite{chen2018strong}, the
electric-magnetic duality
\cite{bae2001electric,seiberg1995electric,hatsuda1999electric}, the duality in
the couple of gauge fields to gravities \cite{igarashi1998electric}, the
duality in higher spin gauge fields \cite{boulanger2003note}, the duality in
quantum many-body systems \cite{zarei2017strong}, the duality in string theory
\cite{duff1994four,font1990strong,alvarez1995introduction,giveon1994target}.
In condensed matter physics, there are also dualities, such as the duality
between the Ising and the Heisenberg models and the gauge theory
\cite{sugamoto1979dual}. Moreover, the theory of the quantum sine-Gordon
equation is equivalent to the theory of massive Thirring model when the
parameter satisfying certain conditions
\cite{PhysRevD.11.2088,PhysRevD.11.3026}; such a duality exists in the lattice
sine-Gordon model \cite{PhysRevD.101.074503}. By this duality, the soliton
solution of the non-linear Schr\"{o}dinger model can be studied by the
Thirring model fermion in the non-relativistic limit \cite{Ghosh2019}.

In section \ref{Duality}, we present the duality relation. In section
\ref{Solving}, we show that the duality relation can serve as a method of
solving field equations. In sections \ref{Power}, \ref{Sine-Gordon}, and
\ref{exp}, as examples, we discuss the duality of the power-interaction field,
the sine-Gordon field, etc. In appendix \ref{CQmechanics}, we sketch a similar
duality in classical mechanics and in quantum mechanics as a comparison. In
appendix \ref{travelingwavesolution}, we calculate a traveling wave solution
of scalar field equation.

\section{The duality transformation and the duality family \label{Duality}}

In the section, we show that there exists a duality transformation between
self-interacting massive scalar fields. All fields connected by the duality
transformation form a family.%

\begin{tcolorbox}[boxrule=0pt,
  boxsep=0pt,
  colback={lightgray1},
  enhanced jigsaw,
  borderline west={3pt}{0pt}{lightgray2},
  sharp corners,
  before skip=10pt,
  after skip=10pt,
breakable,]
\textit{Two massive scalar fields }$\phi\left(  x\right)  $\textit{ and
}$\varphi\left(  y\right)  $\textit{,}%
\begin{align}
\square\phi+m^{2}\phi+\frac{\partial V\left(  \phi\right)  }{\partial\phi} &
=0,\label{phieqa}\\
\square\varphi+M^{2}\varphi+\frac{\partial U\left(  \varphi\right)  }%
{\partial\varphi} &  =0\label{varphieqa1}%
\end{align}
\textit{\ with }$m$\textit{ and }$M$\textit{ the masses, if the potentials
}$V\left(  \phi\right)  $\textit{ and }$U\left(  \varphi\right)  $\textit{ are
related by}%
\begin{align}
\frac{1}{m^{2}}\phi^{-2}\left[  G-V\left(  \phi\right)  \right]   &  =\frac
{1}{M^{2}}\varphi^{-2}\left[  \mathcal{G-}U\left(  \varphi\right)  \right]
,\label{Uphiw}\\
\phi &  \leftrightarrow\varphi^{\sigma},
\end{align}
\textit{where}%
\begin{align}
G &  =\frac{1}{2}\partial_{\mu}\phi\partial^{\mu}\phi+\frac{1}{2}m^{2}\phi
^{2}+V\left(  \phi\right)  ,\label{Ephi}\\
\mathcal{G} &  =\frac{1}{2}\partial_{\mu}\varphi\partial^{\mu}\varphi+\frac
{1}{2}M^{2}\varphi^{2}+U\left(  \varphi\right)  ,\label{GEphi}%
\end{align}
\textit{the fields }$\phi\left(  x\right)  $\textit{ and }$\varphi\left(
y\right)  $\textit{ are related by the duality transformation:}%
\begin{align}
\phi &  \leftrightarrow\varphi^{\sigma},\label{phitans}\\
x^{\mu} &  \leftrightarrow\frac{M}{m}\sigma y^{\mu},\text{ \ }\mu
=0,1,\ldots.\label{xtans}%
\end{align}
\textit{The constant }$\sigma$\textit{ can be chosen arbitrarily.}%

\end{tcolorbox}

The above duality relation can be verified directly.

The duality relation (\ref{Uphiw}) is an explicit expression of the dual
potential. Given a potential $V\left(  \phi\right)  $, the dual potential by
Eq. (\ref{Uphiw}) reads%
\begin{equation}
U\left(  \varphi\right)  =\left.  \frac{M^{2}}{m^{2}}\phi^{-2}\left[  V\left(
\phi\right)  -G\left(  x\right)  \right]  \right\vert _{\substack{\phi
=\varphi^{\sigma}\\x^{\mu}=\frac{M}{m}\sigma y^{\mu}}}\varphi^{2}%
+\mathcal{G}\left(  y\right)  .
\end{equation}

Note that $G$\ defined by Eq. (\ref{Ephi}) does not explicitly depend on
$\phi$, i.e.,
\begin{equation}
\frac{\partial G}{\partial\phi}=0. \label{dGeq0}%
\end{equation}
This can be verified as follows. Without loss of generality, we consider the
$1+1$-dimensional case. In $1+1$ dimensions, Eq. (\ref{Ephi}) becomes
\begin{equation}
G=\frac{1}{2}\left(  \frac{\partial\phi}{\partial t}\right)  ^{2}-\frac{1}%
{2}\left(  \frac{\partial\phi}{\partial x}\right)  ^{2}+\frac{1}{2}m^{2}%
\phi^{2}+V\left(  \phi\right)  \label{G11}%
\end{equation}
and the field equation (\ref{phieqa}) is%
\begin{equation}
\frac{\partial^{2}\phi}{\partial t^{2}}-\frac{\partial^{2}\phi}{\partial
x^{2}}+m^{2}\phi+\frac{\partial V\left(  \phi\right)  }{\partial\phi}=0.
\label{eq11}%
\end{equation}
By Eq. (\ref{G11}) we have
\begin{align}
\frac{\partial G}{\partial\phi}  &  =\frac{\partial\phi}{\partial t}%
\frac{\partial}{\partial\phi}\left(  \frac{\partial\phi}{\partial t}\right)
-\frac{\partial\phi}{\partial x}\frac{\partial}{\partial\phi}\left(
\frac{\partial\phi}{\partial x}\right)  +m^{2}\phi+\frac{\partial V\left(
\phi\right)  }{\partial\phi}\nonumber\\
&  =\frac{\partial\phi}{\partial t}\left[  \frac{\partial t}{\partial\phi
}\frac{\partial}{\partial t}\left(  \frac{\partial\phi}{\partial t}\right)
+\frac{\partial x}{\partial\phi}\frac{\partial}{\partial x}\left(
\frac{\partial\phi}{\partial t}\right)  \right]  -\frac{\partial\phi}{\partial
x}\left[  \frac{\partial t}{\partial\phi}\frac{\partial}{\partial t}\left(
\frac{\partial\phi}{\partial x}\right)  +\frac{\partial x}{\partial\phi}%
\frac{\partial}{\partial x}\left(  \frac{\partial\phi}{\partial x}\right)
\right] \nonumber\\
&  +m^{2}\phi+\frac{\partial V\left(  \phi\right)  }{\partial\phi}\nonumber\\
&  =\frac{\partial^{2}\phi}{\partial t^{2}}-\frac{\partial^{2}\phi}{\partial
x^{2}}+m^{2}\phi+\frac{\partial V\left(  \phi\right)  }{\partial\phi},
\label{dG}%
\end{align}
where $\frac{\partial\phi}{\partial t}=\frac{\partial\left(  \phi,x\right)
}{\partial\left(  t,x\right)  }$, $\frac{\partial\phi}{\partial x}%
=\frac{\partial\left(  \phi,t\right)  }{\partial\left(  x,t\right)  }$,
$\frac{\partial x}{\partial\phi}=\frac{\partial\left(  x,t\right)  }%
{\partial\left(  \phi,t\right)  }$, and $\frac{\partial t}{\partial\phi}%
=\frac{\partial\left(  t,x\right)  }{\partial\left(  \phi,x\right)  }$ with
$\frac{\partial\left(  u,v\right)  }{\partial\left(  x,y\right)  }=\left(
\begin{array}
[c]{cc}%
\frac{\partial u}{\partial x} & \frac{\partial u}{\partial y}\\
\frac{\partial v}{\partial x} & \frac{\partial v}{\partial,y}%
\end{array}
\right)  $ the Jacobian determinant are used. Comparing Eq. (\ref{dG}) with
the field equation (\ref{eq11}) gives Eq. (\ref{dGeq0}).

Generally speaking, $G$ is a function of $t\ $and $x$, i.e., $G=G\left(
t,x\right)  $. We show in appendix\ \ref{travelingwavesolution} that for
traveling wave solutions $G=c$, a constant, while for nontraveling wave
solutions $G=G\left(  t,x\right)  $.

Comparing the field equations (\ref{phieqa}) and (\ref{varphieqa1}) shows that
the duality transformation preserves the type of the field equation: a
Klein-Gordon type equation is still a Klein-Gordon type equation after the
duality transformation.

\textit{Duality family.} In the duality relation, the constant $\sigma$ can be
chosen arbitrarily and different choices give different dual fields, so a
field has an infinite number of dual fields. All the fields who are dual to
each other form a duality family. In a duality family, as long as one equation
is solved, the solutions of other family members can be obtained directly by
the duality transformation, just like that in classical mechanics the solution
of the Newtonian gravitational potential can be obtained from the solution of
the harmonic-oscillator potential by the Newton-Hooke duality
transformation.\qquad\qquad\qquad\qquad\qquad\qquad\qquad\qquad\qquad
\qquad\qquad\qquad\qquad\qquad\qquad\qquad\qquad\qquad\qquad\qquad\qquad
\qquad\qquad\qquad\qquad\qquad\qquad\qquad\qquad\qquad\qquad\qquad\qquad
\qquad\qquad\qquad\qquad\qquad\qquad\qquad\qquad\qquad\qquad\qquad\qquad
\qquad\qquad\qquad\qquad\qquad\qquad\qquad\qquad\qquad\qquad\qquad\qquad
\qquad\qquad\qquad\qquad\qquad\qquad\qquad\qquad\qquad\qquad\qquad\qquad
\qquad\qquad\qquad\qquad\qquad\qquad\qquad\qquad\qquad\qquad\qquad\qquad
\qquad\qquad\qquad\qquad\qquad\qquad\qquad\qquad\qquad\qquad\qquad\qquad
\qquad\qquad\qquad\qquad\qquad\qquad\qquad\qquad\qquad\qquad\qquad\qquad
\qquad\qquad\qquad\qquad\qquad\qquad\qquad\qquad\qquad\qquad\qquad\qquad
\qquad\qquad\qquad\qquad\qquad\qquad\qquad\qquad\qquad\qquad\qquad\qquad
\qquad\qquad\qquad\qquad\qquad\qquad\qquad\qquad\qquad\qquad\qquad\qquad
\qquad\qquad\qquad\qquad\qquad\qquad\qquad\qquad\qquad\qquad\qquad\qquad
\qquad\qquad\qquad\qquad\qquad\qquad\qquad\qquad\qquad\qquad\qquad\qquad
\qquad\qquad\qquad\qquad\qquad\qquad\qquad\qquad\qquad\qquad\qquad\qquad
\qquad\qquad\qquad\qquad\qquad\qquad\qquad\qquad\qquad\qquad\qquad\qquad
\qquad\qquad\qquad\qquad\qquad\qquad\qquad\qquad\qquad\qquad\qquad\qquad
\qquad\qquad\qquad\qquad\qquad\qquad\qquad\qquad\qquad\qquad\qquad\qquad
\qquad\qquad\qquad\qquad\qquad\qquad\qquad\qquad\qquad\qquad\qquad\qquad
\qquad\qquad\qquad\qquad\qquad\qquad\qquad\qquad\qquad\qquad\qquad\qquad\qquad

The duality transformation with $\sigma=1$ is a special duality transformation
which only changes the coupling constant. To see this, we extract the coupling
constant $\lambda$ out of the potential and rewrite the potential $V\left(
\phi\right)  $ as $\lambda V\left(  \phi\right)  $. The dual field of $\lambda
V\left(  \phi\right)  $ is $\frac{M^{2}}{m^{2}}\lambda V\left(  \varphi
\right)  $ (choosing $\frac{M^{2}}{m^{2}}G=\mathcal{G}$). That is, the
coupling constant $\lambda$ is transformed to $\frac{M^{2}}{m^{2}}\lambda$.
This allows us to transform a large-mass strongly interacting field to a
small-mass weakly interacting field which can be dealt with perturbatively.

\section{Solving field equations by the duality \label{Solving}}

In the above, we show that if two fields satisfy the duality relation
(\ref{Uphiw}), the solutions of the field equations (\ref{phieqa}) and
(\ref{varphieqa1}) are related by the duality transformations (\ref{phitans})
and (\ref{xtans}). This provides an approach for solving field equations:
starting from a solved field, we can obtain the solution of its dual field by
the duality transformation.

First we show that the duality relation transforms the solution of a field
equation to the solution of the dual field equation. The field equation
(\ref{phieqa}) has an implicit solution:%
\begin{equation}
\beta_{\mu}x^{\mu}+\int\frac{\sqrt{-\beta^{2}}}{\sqrt{2\left(  \frac{1}%
{2}m^{2}\phi^{2}+V\left(  \phi\right)  -G\right)  }}d\phi=0 \label{solphia}%
\end{equation}
with $\beta^{2}=\beta_{\mu}\beta^{\mu}$ a constant. This is a traveling wave
solution of the scalar field equation, see Appendix
\ref{travelingwavesolution}. Note that for traveling wave solutions $G$ is a constant.

Substituting the duality relations (\ref{phitans}) and (\ref{xtans}) into the
solution (\ref{solphia}),%
\begin{equation}
\beta_{\mu}y^{\mu}+\int\frac{\sqrt{-\beta^{2}}}{\sqrt{2\left(  \frac{1}%
{2}M^{2}\varphi^{2}+\frac{M^{2}}{m^{2}}\varphi^{2\left(  1-\sigma\right)
}\left[  V\left(  \varphi^{\sigma}\right)  -G\right]  \right)  }}d\varphi=0,
\label{solivarphiA1}%
\end{equation}
we arrive at an implicit solution of a field equation with the potential
$U\left(  \varphi\right)  =\frac{M^{2}}{m^{2}}\varphi^{2\left(  1-\sigma
\right)  }\left[  V\left(  \varphi^{\sigma}\right)  -G\right] $ $+\mathcal{G}$:%
\begin{equation}
\beta_{\mu}y^{\mu}+\int\frac{\sqrt{-\beta^{2}}}{\sqrt{2\left(  \frac{1}%
{2}M^{2}\varphi^{2}+U\left(  \varphi\right)  -\mathcal{G}\right)  }}%
d\varphi=0. \label{solivarphi}%
\end{equation}

Next we show that the solutions of a field equation and its dual field
equation are related by the duality relation. Eq. (\ref{Uphiw}) gives%
\begin{equation}
G=V\left(  \phi\right)  -\frac{m^{2}\phi^{2}}{M^{2}\varphi^{2}}\left[
U\left(  \varphi\right)  -\mathcal{G}\right]  . \label{Gg}%
\end{equation}
Rewriting the solution of the potential $V\left(  \phi\right)  $, Eq.
(\ref{solphia}), as%
\begin{equation}
\beta_{\mu}\frac{dx^{\mu}}{d\phi}+\frac{\sqrt{-\beta^{2}}}{\sqrt{2\left(
\frac{1}{2}m^{2}\phi^{2}+V\left(  \phi\right)  -G\right)  }}=0 \label{vphi1}%
\end{equation}
and substituting Eq. (\ref{Gg}) into Eq. (\ref{vphi1}) give%
\begin{equation}
\beta_{\mu}\frac{m\phi}{M\varphi}\frac{dx^{\mu}}{d\phi}+\frac{\sqrt{-\beta
^{2}}}{\sqrt{2\left(  \frac{1}{2}M^{2}\varphi^{2}+U\left(  \varphi\right)
-\mathcal{G}\right)  }}=0 \label{afterT}%
\end{equation}
Eq. (\ref{afterT}) should be a solution of the field equation with the
potential $U\left(  \varphi\right)  $, i.e., it must be of the form%
\begin{equation}
\beta_{\mu}\frac{dy^{\mu}}{d\varphi}+\frac{\sqrt{-\beta^{2}}}{\sqrt{2\left(
\frac{1}{2}M^{2}\varphi^{2}+U\left(  \varphi\right)  -\mathcal{G}\right)  }%
}=0. \label{afterTmust}%
\end{equation}
Comparing Eqs. (\ref{afterT}) and (\ref{afterTmust}) gives%
\begin{equation}
\frac{m\phi}{M\varphi}\frac{dx^{\mu}}{d\phi}=\frac{dy^{\mu}}{d\varphi}.
\end{equation}
\qquad We have
\begin{equation}
\frac{m}{M}\frac{dx^{\mu}}{dy^{\mu}}=\frac{d\ln\phi}{d\ln\varphi}=\sigma,
\label{Eqphixy1}%
\end{equation}
where $\sigma$ is an arbitrary constant. Solving Eq. (\ref{Eqphixy1}) gives
the duality transformations (\ref{phitans}) and (\ref{xtans}).

\section{Power interactions \label{Power}}

The duality of a power interaction $V\left(  \phi\right)  =\lambda\phi^{a}$,
generally, is no longer a power interaction. If requiring that duality of a
power interaction is still a power interaction, we arrive at the following conclusion.

The dual potential of the power potential%
\begin{equation}
V\left(  \phi\right)  =\lambda\phi^{a},
\end{equation}
by Eq. (\ref{Uphiw}), is
\begin{equation}
U\left(  \varphi\right)  =\frac{M^{2}}{m^{2}}\lambda\varphi^{2+\left(
a-2\right)  \sigma}-\frac{M^{2}}{m^{2}}G\varphi^{2\left(  1-\sigma\right)
}+\mathcal{G}. \label{dualPower}%
\end{equation}
If requiring the dual potential is still a power potential, i.e.,%
\begin{equation}
U\left(  \varphi\right)  =\eta\varphi^{A},
\end{equation}
there are two choices. One choice is $2+\left(  a-2\right)  \sigma=0$ and
$A=2\left(  1-\sigma\right)  $, i.e.,%
\begin{equation}
\sigma=\frac{2}{2-a},\text{ \ }\frac{2}{2-a}=\frac{2-A}{2}.
\end{equation}
This gives $U\left(  \varphi\right)  =-\frac{M^{2}}{m^{2}}G\varphi^{\frac
{2a}{a-2}}+\frac{M^{2}}{m^{2}}\lambda+\mathcal{G}$. Choosing $\mathcal{G}%
=-\frac{M^{2}}{m^{2}}\lambda$ so that the dual potential is still a power
potential, we arrive at%
\begin{equation}
U\left(  \varphi\right)  =-\frac{M^{2}}{m^{2}}G\varphi^{\frac{2a}{a-2}}.
\end{equation}

Another choice is $A=2+(a-2)\sigma$ and $2\left(  1-\sigma\right)  =0$, i.e.,%
\begin{equation}
\sigma=1,\text{ \ }A=a.
\end{equation}
This gives $U\left(  \varphi\right)  =\frac{M^{2}}{m^{2}}\lambda\varphi
^{a}-\frac{M^{2}}{m^{2}}G+\mathcal{G}$. Choosing $\mathcal{G=}-\frac{M^{2}%
}{m^{2}}G$ so that the dual potential is still a power potential, we arrive at%
\begin{equation}
U\left(  \varphi\right)  =\frac{M^{2}}{m^{2}}\lambda\varphi^{a}.
\end{equation}

In this case, the power of the potential does not change after the duality
transformation, but the coupling constant becomes $\frac{M^{2}}{m^{2}}\lambda
$. That is, the duality transformation of $\sigma=1$ transforms a field of
mass $m$ and coupling constant $\lambda$ into a field of mass $M$ and coupling
constant $\frac{M^{2}}{m^{2}}\lambda$. When $M\ll m$, the coupling constant of
the dual field $\frac{M^{2}}{m^{2}}\lambda\ll\lambda$. In this case, the
duality connects a large-mass and strong-coupling field to a small-mass and
weak-coupling field.

Especially, when $M=m$ the duality transformation with $\sigma=1$ is an
identity transformation.

\subsection{The $\phi^{4}$-field: self-duality}

The $\phi^{4}$-field is self-dual. By \textit{self-dual} we mean\ that the
duality of a $\phi^{4}$-field is also a $\phi^{4}$-field. That is, the fields
\begin{equation}
V\left(  \phi\right)  =\lambda\phi^{4}\text{ \ and }U\left(  \varphi\right)
=\eta\varphi^{4}%
\end{equation}
are dual. The duality transformations are%
\begin{equation}
\phi\rightarrow\varphi^{-1},\text{ }x^{\mu}\rightarrow-\frac{M}{m}y^{\mu
},\text{ }\frac{m^{2}}{M^{2}}\eta\rightarrow-G. \label{dualphi4}%
\end{equation}

It can be checked that the field equation with $V\left(  \phi\right)
=\lambda\phi^{4}$,%
\begin{equation}
\square\phi+m^{2}\phi+4\lambda\phi^{3}=0, \label{eqph4}%
\end{equation}
has a soliton solution%
\begin{equation}
\phi=\frac{m}{2\sqrt{\lambda}}\tan\left(  \alpha t+\beta x_{1}+\gamma
x_{2}-x_{3}\sqrt{\alpha^{2}-\beta^{2}-\gamma^{2}+\frac{m^{2}}{2}}%
+\delta\right)  . \label{phi4self}%
\end{equation}
The solution (\ref{phi4self}) gives $G=\frac{1}{2}\partial_{\mu}\phi
\partial^{\mu}\phi+\frac{m^{2}}{2}\phi^{2}+\lambda\phi^{4}=-\frac{m^{4}%
}{16\lambda}$. The Lorentz invariance of the solution can be verified by
directly performing the Lorentz transformation.

By the duality relation, the solution of the field equation with the potential
$U\left(  \varphi\right)  =\eta\varphi^{4}$,%
\begin{equation}
\square\varphi+M^{2}\varphi+4\eta\varphi^{3}=0,
\end{equation}
is%
\begin{equation}
\varphi=\frac{M}{2\sqrt{\eta}}\tan\left(  \alpha^{\prime}\tau+\beta^{\prime
}y_{1}+\gamma^{\prime}y_{2}-y_{3}\sqrt{\alpha^{\prime2}-\beta^{\prime2}%
-\gamma^{\prime2}+\frac{M^{2}}{2}}+\delta^{\prime}\right)  ,
\end{equation}
where $\alpha^{\prime}=\alpha\frac{M}{m}$,\ $\beta^{\prime}=\beta\frac{M}{m}$,
$\gamma^{\prime}=\gamma\frac{M}{m}$, and $\delta^{\prime}=\delta+\frac{\pi}%
{2}$. After the duality transformation, the mass $m$ is transformed to $M$.

The $\phi^{4}$-field is self-dual, i.e., the duality of a $\phi^{4}$-field is
still a $\phi^{4}$-field. It should be emphasized that even in the case that
the masses of the field and its dual field are the same, the duality
transformation ($\sigma=-1$) is not an identity transformation ($\sigma=1$).

Now we consider the duality between a weak coupling field and a strong
coupling field, taking the $\phi^{4}$-field as an example.

The dual field of a $\phi^{4}$-field with mass $m$ and coupling constant
$\lambda$, by the duality transformation (\ref{dualphi4}), is a $\phi^{4}%
$-field with mass $M$ and coupling constant $\eta=-\frac{M^{2}}{m^{2}}%
G=\frac{m^{2}M^{2}}{16\lambda}$. If the mass of the dual field $\varphi$ is
chosen as $M=\frac{1}{m}$, then the coupling constant becomes%
\begin{equation}
\eta=\frac{1}{16\lambda}.
\end{equation}
The field equation of the dual field is%
\begin{equation}
\square\varphi+\frac{1}{m^{2}}\varphi+\frac{1}{4\lambda}\varphi^{3}=0.
\label{dualeq}%
\end{equation}

This means that the dual field of a $\phi^{4}$-field with mass $m$ and
coupling constant $\lambda$ is a $\phi^{4}$-field with mass $\frac{1}{m}$ and
coupling constant $\frac{1}{16\lambda}$. When the coupling constant $\lambda$
of the field $\phi$ is large (strongly interacting), the coupling constant
$\frac{1}{16\lambda}$ of its dual field $\varphi$ will be small (weakly
interacting). That is, the duality in this case is a strong-weak coupling
duality. In principle, this strong-weak coupling duality also exists in more
general cases, not limited to $\phi^{4}$-field. Especially, if $m$, the mass
of the field $\phi$, is small, then $\frac{1}{m^{2}}$, the mass of the dual
field $\varphi$, is large. This implies that one can\ construct an effective
theory for the field $\varphi$ with a heavy mass propagator, like that in the
four-fermion effective theory.

\subsection{The $\phi^{1}$-field and the $\phi^{-2}$-field}

The dual field of\ the $\phi^{1}$-field is the $\phi^{-2}$-field, i.e., the
fields
\begin{equation}
V\left(  \phi\right)  =\lambda\phi\text{ \ and }U\left(  \varphi\right)
=\eta\varphi^{-2} \label{Uphiwm2}%
\end{equation}
are dual to each other. The duality transformations are%
\begin{equation}
\phi\rightarrow\varphi^{2},\text{ }x^{\mu}\rightarrow2\frac{M}{m}y^{\mu
},\text{ }\frac{m^{2}}{M^{2}}\eta\rightarrow-G.
\end{equation}

It can be checked that the field equation with $V\left(  \phi\right)
=\lambda\phi$ has a $1+3$-dimensional traveling wave solution%
\begin{equation}
\phi=\exp\left(  \alpha t+\beta x_{1}+\gamma x_{2}+\sqrt{m^{2}+\alpha
^{2}-\beta^{2}-\gamma^{2}}x_{3}\right)  -\frac{\lambda}{m^{2}}.
\label{phi4dsol}%
\end{equation}
The solution (\ref{phi4dsol}) gives $G=\frac{1}{2}\partial_{\mu}\phi
\partial^{\mu}\phi+\frac{1}{2}m^{2}\phi^{2}+\lambda\phi=-\frac{\lambda^{2}%
}{2m^{2}}$. For the traveling wave solution, $G$ is a constant.

By the duality transformation, the solution of the $\phi^{-2}$-field is
\begin{equation}
\varphi=\left[  \exp\left(  2\left(  \alpha^{\prime}\tau+\beta^{\prime}%
y_{1}+\gamma^{\prime}y_{2}+\sqrt{M^{2}+\alpha^{\prime2}-\beta^{\prime2}%
-\gamma^{\prime2}}y_{3}\right)  \right)  -\frac{\sqrt{2\eta}}{M}\right]
^{1/2},
\end{equation}
where $\alpha^{\prime}=\alpha\frac{M}{m}$,\ $\beta^{\prime}=\beta\frac{M}{m}$,
and $\gamma^{\prime}=\gamma\frac{M}{m}$.

Moreover, $\phi^{1}$-field has a $1+1$-dimensional nontraveling wave solution%
\begin{equation}
\phi=e^{\alpha t}\sinh\left(  \sqrt{\alpha^{2}+m^{2}}x\right)  -\frac{\lambda
}{m^{2}}. \label{phi1}%
\end{equation}
For this nontraveling wave solution, $G=-\frac{\lambda^{2}}{2m^{2}}%
-\frac{\alpha^{2}+m^{2}}{2}e^{2\alpha t}$ is not a constant but depends on
$t$. This gives $\eta=-\frac{M^{2}}{m^{2}}G=\frac{\lambda^{2}M^{2}}{2m^{4}%
}+\frac{\alpha^{2}+m^{2}}{2m^{2}}M^{2}e^{2\alpha t}$. By the duality
transformation, the corresponding solution of the $\phi^{-2}$-field is%
\begin{equation}
\varphi=\left[  e^{2\frac{M}{m}\alpha\tau}\sinh\left(  2\frac{M}{m}%
\sqrt{\alpha^{2}+M^{2}}y\right)  -\frac{\lambda}{m^{2}}\right]  ^{1/2}.
\end{equation}
This is just the solution of the field equation (\ref{varphieqa1}) with the
potential
\begin{equation}
U\left(  \varphi\right)  =\left(  \frac{\lambda^{2}M^{2}}{2m^{4}}+\frac
{\alpha^{2}+m^{2}}{2m^{2}}M^{2}e^{4\frac{M}{m}\alpha\tau}\right)  \varphi
^{-2}. \label{Uphit}%
\end{equation}
This potential is a $\phi^{-2}$-potential with a time-dependent coefficient.

In a word, the function $G$ for traveling wave solutions is a constant, and
for nontraveling wave solution is not a constant (see Appendix
\ref{travelingwavesolution}).

\subsection{The $\phi^{3}$-field and the $\phi^{6}$-field}

The dual field of\ a $\phi^{3}$-field is a $\phi^{6}$-field, i.e., the fields
\begin{equation}
V\left(  \phi\right)  =\lambda\phi^{3}\text{ \ and \ }U\left(  \varphi\right)
=\eta\phi^{6} \label{Uphi6}%
\end{equation}
are dual to each other. The duality transformations are%
\begin{equation}
\phi\rightarrow\varphi^{-2},\text{ }x^{\mu}\rightarrow-2\frac{M}{m}y^{\mu
},\text{ }\frac{m^{2}}{M^{2}}\eta\rightarrow-G. \label{Txy36}%
\end{equation}

It can be checked that the field equation with $V\left(  \phi\right)
=\lambda\phi^{3}$ has a solution%
\begin{equation}
\phi=-\frac{m^{2}}{6\lambda}\left[  3\tanh^{2}\left(  \alpha x_{1}+\beta
x_{2}+\gamma x_{3}+\sqrt{\alpha^{2}+\beta^{2}+\gamma^{2}+\left(  \frac{m}%
{2}\right)  ^{2}}t\right)  -1\right]  . \label{phi34dsol}%
\end{equation}
The solution (\ref{phi34dsol}) gives $G=\frac{1}{2}\partial_{\mu}\phi
\partial^{\mu}\phi+\frac{1}{2}m^{2}\phi^{2}+\lambda\phi^{3}=\frac{m^{6}%
}{54\lambda^{2}}$.

By the duality transformation, the solution of the $\phi^{6}$-field is%
\begin{equation}
\varphi=\left\{  \frac{\sqrt{-6\eta}}{2M}\left[  3\tanh^{2}\left(  2\left(
\alpha^{\prime}y_{1}+\beta^{\prime}y_{2}+\gamma^{\prime}y_{3}+\sqrt
{\alpha^{\prime2}+\beta^{\prime2}+\gamma^{\prime2}+\left(  \frac{M}{2}\right)
^{2}}\tau\right)  \right)  -1\right]  \right\}  ^{-1/2},
\end{equation}
where $\alpha^{\prime}=\alpha\frac{M}{m}$,\ $\beta^{\prime}=\beta\frac{M}{m}%
$,\ and $\gamma^{\prime}=\gamma\frac{M}{m}$.

\section{The sine-Gordon equation, the sinh-Gordon equation, and all that
\label{Sine-Gordon}}

The sine-Gordon equation, sinh-Gordon equation, and all field equations of
this type can be compactly written as%
\begin{equation}
\square\phi-ae^{\beta\phi}+be^{-\beta\phi}=0 \label{doubexp}%
\end{equation}
which recovers the sine-Gordon equation when $a=b=-\frac{1}{2i}\frac{m^{3}%
}{\sqrt{\lambda}}$ and $\beta=i\frac{\sqrt{\lambda}}{m}$, recovers the
sinh-Gordon equation when $a=b=-\frac{1}{2}$ and $\beta=1$, and so on.

The potential corresponding to the field equation (\ref{doubexp}) is
\begin{equation}
V\left(  \phi\right)  =-\frac{a}{\beta}e^{\beta\phi}-\frac{b}{\beta}%
e^{-\beta\phi}. \label{doubexpV}%
\end{equation}
The field equation (\ref{doubexp}) has the following solution
\cite{polyanin2016handbook}:%
\begin{equation}
\phi\left(  t,x\right)  =\frac{4}{\beta}\operatorname*{arctanh}\left(
\exp\left(  \sqrt{\frac{2\beta\sqrt{ab}}{\mu^{2}-\nu^{2}}}\left(  \mu t+\nu
x+\theta\right)  \right)  \right)  +\frac{1}{2\beta}\ln\frac{b}{a}.
\label{solofdoubexpV}%
\end{equation}
For the potential (\ref{doubexpV}), by Eqs. (\ref{Ephi}) and
(\ref{solofdoubexpV}), we have $G=-2\sqrt{ab}/\beta$.

The dual potential of\ the potential (\ref{doubexpV}) can be obtained by the
duality relations (\ref{Uphiw}). For massless fields, $m^{2}/M^{2}=1$. The
function $\mathcal{G}\left(  y\right)  $ can be chosen arbitrarily, because
the choice of $\mathcal{G}\left(  y\right)  $ does not influence the field
equation. Choosing $\mathcal{G}\left(  y\right)  =0$ gives%
\begin{equation}
U\left(  \varphi\right)  =\frac{1}{\beta}\varphi^{2\left(  1-\sigma\right)
}\left[  2\sqrt{ab}-\left(  ae^{\beta\varphi^{\sigma}}+be^{-\beta
\varphi^{\sigma}}\right)  \right]  . \label{doublexp}%
\end{equation}
Different choices of the constant $\sigma$ in Eq. (\ref{doublexp}) give
different dual potentials.

The dual field equation by the duality transformations (\ref{phitans}) and
(\ref{xtans}) is%
\begin{equation}
\square\varphi-\frac{2\left(  1-\sigma\right)  }{\beta}\varphi^{1-2\sigma
}\left(  ae^{\beta\varphi^{\sigma}}+be^{-\beta\varphi^{\sigma}}-2\sqrt
{ab}\right)  -\sigma\varphi^{1-\sigma}\left(  ae^{\beta\varphi^{\sigma}%
}-be^{-\beta\varphi^{\sigma}}\right)  =0,
\end{equation}
and the solution of the dual field,%
\begin{equation}
\varphi\left(  \tau,y\right)  =\left[  \frac{4}{\beta}\operatorname*{arctanh}%
\left(  \exp\left(  \sqrt{\frac{2\beta\sqrt{ab}}{\mu^{2}-\nu^{2}}}\left(
\mu\sigma\tau+\nu\sigma y+\theta\right)  \right)  \right)  +\frac{1}{2\beta
}\ln\frac{b}{a}\right]  ^{1/\sigma}.
\end{equation}

Note that different solutions lead to different $G$ and then leads to
different coefficients in the dual potential.

\subsection{The sine-Gordon equation}

The field equation (\ref{doubexp}) recovers the sine-Gordon equation when
$a=b=-\frac{1}{2i}\frac{m^{3}}{\sqrt{\lambda}}$ and $\beta=i\frac
{\sqrt{\lambda}}{m}$:%
\begin{equation}
\square\phi+\frac{m^{3}}{\sqrt{\lambda}}\sin\frac{\sqrt{\lambda}}{m}\phi=0.
\label{sinG}%
\end{equation}
The potential is
\begin{equation}
V\left(  \phi\right)  =2\frac{m^{4}}{\lambda}\sin^{2}\left(  \frac
{\sqrt{\lambda}}{2m}\phi\right)  .
\end{equation}
The solution of the sine-Gordon equation by Eq. (\ref{solofdoubexpV}) is%
\begin{equation}
\phi\left(  t,x\right)  =\frac{4m}{\sqrt{\lambda}}\arctan\left(  \exp\left(
\frac{m}{\sqrt{\nu^{2}-\mu^{2}}}\left(  \mu t+\nu x+\theta\right)  \right)
\right)  . \label{solutionsG}%
\end{equation}
For the sine-Gordon field $G=-2\sqrt{ab}/\beta=-m^{4}/\lambda$. The dual
potential of the sine-Gordon potential then is%
\begin{equation}
U\left(  \varphi\right)  =2\frac{m^{4}}{\lambda}\varphi^{2\left(
1-\sigma\right)  }\sin^{2}\left(  \frac{\sqrt{\lambda}}{2m}\varphi^{\sigma
}\right)  .
\end{equation}
The dual equation is%
\begin{equation}
\square\varphi+4\left(  1-\sigma\right)  \frac{m^{4}}{\lambda}\varphi
^{1-2\sigma}\sin^{2}\left(  \frac{\sqrt{\lambda}}{2m}\varphi^{\sigma}\right)
+\sigma\frac{m^{3}}{\sqrt{\lambda}}\varphi^{1-\sigma}\sin\left(  \frac
{\sqrt{\lambda}}{m}\varphi^{\sigma}\right)  =0. \label{sin1}%
\end{equation}
The solution of the dual field equation (\ref{sin1}) is%
\begin{equation}
\varphi\left(  \tau,y\right)  =\left[  \frac{4m}{\sqrt{\lambda}}\arctan\left(
\exp\left(  \frac{m}{\sqrt{\nu^{2}-\mu^{2}}}\left(  \sigma\mu\tau+\sigma\nu
y+\theta\right)  \right)  \right)  \right]  ^{1/\sigma}.
\end{equation}

Moreover, the parameter $m$ in the sine-Gordon equation (\ref{sinG}) can be
explained as the mass of the field, which can be seen by expanding Eq.
(\ref{sinG}). If regarding the parameter $m$ as the mass, the potential then
is%
\begin{equation}
V\left(  \phi\right)  =2\frac{m^{4}}{\lambda}\sin^{2}\left(  \frac
{\sqrt{\lambda}}{2m}\phi\right)  -\frac{1}{2}m^{2}\phi^{2}.
\end{equation}
The dual potential of the sine-Gordon potential then is%
\begin{equation}
U\left(  \varphi\right)  =2\frac{M^{4}}{\lambda}\varphi^{2\left(
1-\sigma\right)  }\sin^{2}\left(  \frac{\sqrt{\lambda}}{2M}\varphi^{\sigma
}\right)  -\frac{1}{2}M^{2}\varphi^{2}.
\end{equation}
The dual equation is%
\begin{equation}
\square\varphi+4\left(  1-\sigma\right)  \frac{M^{4}}{\lambda}\varphi
^{1-2\sigma}\sin^{2}\left(  \frac{\sqrt{\lambda}}{2M}\varphi^{\sigma}\right)
+\sigma\frac{M^{3}}{\sqrt{\lambda}}\varphi^{1-\sigma}\sin\left(  \frac
{\sqrt{\lambda}}{M}\varphi^{\sigma}\right)  =0. \label{sin}%
\end{equation}
The solution of the dual field equation (\ref{sin}) is%
\begin{equation}
\varphi\left(  \tau,y\right)  =\left[  \frac{4M}{\sqrt{\lambda}}\arctan\left(
\exp\left(  \frac{M}{\sqrt{\nu^{2}-\mu^{2}}}\left(  \sigma\mu\tau+\sigma\nu
y+\theta\right)  \right)  \right)  \right]  ^{1/\sigma}.
\end{equation}

\subsection{The sinh-Gordon equation}

The field equation (\ref{doubexp}) recovers the sinh-Gordon equation when
$a=b=-\frac{1}{2}$ and $\beta=1$:%
\begin{equation}
\square\phi+\sinh\phi=0.
\end{equation}
The potential is%
\begin{equation}
V\left(  \phi\right)  =2\sinh^{2}\left(  \frac{\phi}{2}\right)  .
\end{equation}
The solution of the sinh-Gordon equation by Eq. (\ref{solofdoubexpV}) is%
\begin{equation}
\phi\left(  t,x\right)  =2\ln\left(  \coth\frac{\mu t+\nu x+\theta}{2\sqrt
{\nu^{2}-\mu^{2}}}\right)  .
\end{equation}
For the sinh-Gordon field, by Eq. (\ref{doublexp}), we have $G=1$. The dual
potential of the sinh-Gordon potential then is%
\begin{equation}
U\left(  \varphi\right)  =2\varphi^{2\left(  1-\sigma\right)  }\sinh
^{2}\left(  \frac{\varphi^{\sigma}}{2}\right)  .
\end{equation}
The dual equation is%
\begin{equation}
\square\varphi+4\left(  1-\sigma\right)  \varphi^{1-2\sigma}\sinh^{2}\left(
\frac{\varphi^{\sigma}}{2}\right)  +\sigma\varphi^{1-\sigma}\sinh\left(
\varphi^{\sigma}\right)  =0. \label{sinh}%
\end{equation}
The solution of the dual field equation (\ref{sinh}) is%
\begin{equation}
\varphi\left(  \tau,y\right)  =\left[  2\ln\left(  \coth\frac{\sigma\mu
\tau+\sigma\nu y+\theta}{2\sqrt{\nu^{2}-\mu^{2}}}\right)  \right]  ^{1/\sigma
}.
\end{equation}

\section{$\square\phi-ae^{\beta\phi}-be^{2\beta\phi}=0$ \label{exp}}

Consider the scalar field equation%
\begin{equation}
\square\phi-ae^{\beta\phi}-be^{2\beta\phi}=0. \label{expbexp2b}%
\end{equation}
The potential is%
\begin{equation}
V\left(  \phi\right)  =-\frac{a}{\beta}e^{\beta\phi}-\frac{b}{2\beta}%
e^{2\beta\phi}. \label{expbexp2bV}%
\end{equation}
It can be checked that the field equation (\ref{expbexp2b}) has the solution:%
\begin{equation}
\phi\left(  t,x\right)  =-\frac{1}{\beta}\ln\left(  \frac{a\beta}{\mu^{2}%
-\nu^{2}}\left[  1+\sqrt{1+\frac{b}{a^{2}\beta}\left(  \mu^{2}-\nu^{2}\right)
}\sin\left(  \mu t+\nu x+\theta\right)  \right]  \right)  .
\label{solofexpbexp2b}%
\end{equation}
For the field $\phi$ with the potential (\ref{expbexp2bV}), we have $G=\left(
\nu^{2}-\mu^{2}\right)  /\left(  2\beta^{2}\right)  $.

The dual potential then by Eq. (\ref{Uphiw}) is%
\begin{equation}
U\left(  \varphi\right)  =\varphi^{2\left(  1-\sigma\right)  }\left(
-\frac{a}{\beta}e^{\beta\varphi^{\sigma}}-\frac{b}{2\beta}e^{2\beta
\varphi^{\sigma}}+\frac{\mu^{2}-\nu^{2}}{2\beta^{2}}\right)  .
\label{expbexp2bdualU}%
\end{equation}
Here $M^{2}/m^{2}=1$ for the field is massless and we choose $\mathcal{G}=0$.

By the duality transformations (\ref{phitans}) and (\ref{xtans}), the dual
field equation is
\begin{equation}
\square\varphi-2\left(  1-\sigma\right)  \varphi^{1-2\sigma}\left(  \frac
{a}{\beta}e^{\beta\varphi^{\sigma}}+\frac{b}{2\beta}e^{2\beta\varphi^{\sigma}%
}-\frac{\mu^{2}-\nu^{2}}{2\beta^{2}}\right)  -\sigma\varphi^{1-\sigma}%
e^{\beta\varphi^{\sigma}}\left(  a+be^{\beta\varphi^{\sigma}}\right)  =0
\end{equation}
and the solution of the dual field equation is%
\begin{equation}
\varphi\left(  \tau,y\right)  =\left\{  -\frac{1}{\beta}\ln\left(
\frac{a\beta}{\mu^{2}-\nu^{2}}\left[  1+\sqrt{1+\frac{b}{a^{2}\beta}\left(
\mu^{2}-\nu^{2}\right)  }\sin\left(  {\sigma}\mu\tau+{\sigma}\nu
y+\theta\right)  \right]  \right)  \right\}  ^{1/\sigma}.
\end{equation}

For more examples see Ref. \cite{li2019duality}.

\section{Conclusion and outlooks}

We show that there exits a duality of scalar fields. The duality
transformation preserves the type of the field equation.

A field has an infinite number of dual fields. All dual fields form a duality
family. In a duality family, as long as one field is solved, all other fields
can be solved by the duality relation. This provides a high-efficiency
approach to solve field equations.

The existence of the duality family inspires us to classify fields based on
the duality. A duality family is a duality class. In future works, we will
discuss the property of the duality family.

The duality relation also relates various qualities of fields, such as heat
kernels, effective actions, vacuum energies, spectral counting functions, etc.
In further works, we will consider the quantum theory of the duality, such as
the duality relation of the Feynman rule. Especially, in quantum field theory
we will consider the duality in the heat kernel method
\cite{vassilevich2003heat} and in the scattering spectrum method
\cite{graham2009spectral,pang2012relation,li2015heat}. In these methods we can
calculate the one-loop effective action and the vacuum
energy\ \cite{dai2009number,dai2010approach}. We may observe the relation of
the one-loop effective action and the vacuum energy of dual fields. A similar
duality also appears in the Gross--Pitaevskii equation \cite{liu2021exactly}.
Moreover, we will consider the duality of spinor fields and vector fields.

\appendix
\setcounter{section}{0}
\numberwithin{equation}{section}
\section{The duality in classical and quantum mechanics
\label{CQmechanics}}

In the above, we discuss the duality of scalar fields. In this appendix, we
show that a similar duality also exists in classical and quantum mechanics.

\subsection{The duality in classical mechanics \label{classical}}

The equation of motion in classical mechanics is the Newton equation.%

\begin{tcolorbox}[boxrule=0pt,
  boxsep=0pt,
  colback={lightgray1},
  enhanced jigsaw,
  borderline west={3pt}{0pt}{lightgray2},
  sharp corners,
  before skip=10pt,
  after skip=10pt,
breakable,]
\textit{Two equations of motion with the one-dimensional potentials }$U\left(
x\right)  $\textit{ and }$V\left(  \xi\right)  $%
\begin{align}
\frac{dt}{dx} &  =\frac{1}{\sqrt{2\left[  E-U\left(  x\right)  \right]  }%
},\label{Ueq1d}\\
\frac{d\tau}{d\xi} &  =\frac{1}{\sqrt{2\left[  \mathcal{E}-V\left(
\xi\right)  \right]  }},\label{Veq1d}%
\end{align}
\textit{where }$E$\textit{ and }$E$\textit{\ are energies, if the potentials
}$U\left(  x\right)  $\textit{ and }$V\left(  \xi\right)  $\textit{ are
related by }%
\begin{equation}
x^{-2}\left[  E-U\left(  x\right)  \right]  =\xi^{-2}\left[  \mathcal{E}%
-V\left(  \xi\right)  \right]  \label{DualoneD}%
\end{equation}
\textit{with}%
\begin{equation}
x\leftrightarrow\xi^{\sigma},\label{rrho1d}%
\end{equation}
\textit{the solutions of the equations of motion (\ref{Ueq1d}) and
(\ref{Veq1d}) are related by the transformation}%
\begin{equation}
t\leftrightarrow\sigma\tau.\label{thetaphi1d}%
\end{equation}
\textit{Here }$\sigma$\textit{ is a constant chosen arbitrarily.}%
\end{tcolorbox}

\textit{Proof. }By\textit{ }Eqs. (\ref{Ueq1d}) and (\ref{Veq1d}) we have
\begin{align}
E-U\left(  x\right)   &  =\frac{1}{2}\left(  \frac{dx}{dt}\right)  ^{2},\\
\mathcal{E}-V\left(  \xi\right)   &  =\frac{1}{2}\left(  \frac{d\xi}{d\tau
}\right)  ^{2}.
\end{align}
Substituting into Eq. (\ref{DualoneD}) gives%
\begin{equation}
x^{-2}\left[  \frac{1}{2}\left(  \frac{dx}{dt}\right)  ^{2}\right]  =\xi
^{-2}\left[  \frac{1}{2}\left(  \frac{d\xi}{d\tau}\right)  ^{2}\right]  .
\end{equation}
This gives%
\begin{equation}
\frac{dt}{d\tau}=\frac{d\ln x}{d\ln\xi}.
\end{equation}
Because $t$ and $\tau$ are independent of $x$ and $\xi$, we have
\begin{equation}
\frac{dt}{d\tau}=\frac{d\ln x}{d\ln\xi}=\sigma,\label{dtdtao}%
\end{equation}
where $\sigma$ is an arbitrary constant. Solving Eq. (\ref{dtdtao}) gives the
duality transformations (\ref{rrho1d}) and (\ref{thetaphi1d}).%

\begin{tcolorbox}[boxrule=0pt,
  boxsep=0pt,
  colback={lightgray1},
  enhanced jigsaw,
  borderline west={3pt}{0pt}{lightgray2},
  sharp corners,
  before skip=10pt,
  after skip=10pt,
breakable,]

\textit{Two orbit equations with the three-dimensional central potentials
}$U\left(  r\right)  $\textit{ and }$V\left(  \rho\right)  $\textit{,}%
\begin{align}
\frac{d\theta}{dr} &  =\frac{l/r^{2}}{\sqrt{2\left[  E-l^{2}/\left(
2r^{2}\right)  -U\left(  r\right)  \right]  }},\label{Ueq}\\
\frac{d\phi}{d\rho} &  =\frac{\ell/\rho^{2}}{\sqrt{2\left[  \mathcal{E}%
-\ell^{2}/\left(  2\rho^{2}\right)  -V\left(  \rho\right)  \right]  }%
},\label{Veq}%
\end{align}
\textit{where }$E$\textit{ and }$E$\textit{\ are energies and }$l$\textit{ and
}$\ell$\textit{ are angular momenta, if the potentials }$U\left(  r\right)
$\textit{ and }$V\left(  \rho\right)  $\textit{ are related by }%
\begin{equation}
\frac{r^{2}}{l^{2}}\left[  E-U\left(  r\right)  \right]  =\frac{\rho^{2}}%
{\ell^{2}}\left[  \mathcal{E-}V\left(  \rho\right)  \right]  \label{rUrhoV}%
\end{equation}
\textit{with}%
\begin{equation}
r\leftrightarrow\rho^{\frac{l}{\ell}\sigma},\label{rrho}%
\end{equation}
\textit{the solutions of the equations of motion (\ref{Ueq}) and (\ref{Veq})
are related by the transformation}%
\begin{equation}
\theta\leftrightarrow\frac{l}{\ell}\sigma\phi.\label{thetaphi}%
\end{equation}
\textit{Here }$\sigma$\textit{ is a constant chosen arbitrarily.}%

\end{tcolorbox}

\textit{Proof. }By Eqs. (\ref{Ueq}) and (\ref{Veq}) we have%
\begin{align}
E-U\left(  r\right)   &  =\frac{1}{2}\frac{l^{2}}{r^{4}}\left(  \frac
{dr}{d\theta}\right)  ^{2}+\frac{l^{2}}{2r^{2}},\\
\mathcal{E-}V\left(  \rho\right)   &  =\frac{1}{2}\frac{\ell^{2}}{\rho^{4}%
}\left(  \frac{d\rho}{d\phi}\right)  ^{2}+\frac{\ell^{2}}{2\rho^{2}}.
\end{align}
Substituting into Eq. (\ref{rUrhoV}) gives
\begin{equation}
\frac{r^{2}}{l^{2}}\left[  \frac{1}{2}\frac{l^{2}}{r^{4}}\left(  \frac
{dr}{d\theta}\right)  ^{2}+\frac{l^{2}}{2r^{2}}\right]  =\frac{\rho^{2}}%
{\ell^{2}}\left[  \frac{1}{2}\frac{\ell^{2}}{\rho^{4}}\left(  \frac{d\rho
}{d\phi}\right)  ^{2}+\frac{\ell^{2}}{2\rho^{2}}\right]  .
\end{equation}
This gives%
\begin{equation}
\left(  \frac{d\ln r}{d\theta}\right)  ^{2}=\left(  \frac{d\ln\rho}{d\phi
}\right)  ^{2},
\end{equation}
\qquad or,%
\begin{equation}
\frac{d\theta}{d\phi}=\frac{d\ln r}{d\ln\rho}.
\end{equation}
Because $\theta$ and $\phi$ are independent of $r$ and $\rho$, we have%
\begin{equation}
\frac{d\theta}{d\phi}=\frac{d\ln r}{d\ln\rho}=\sigma\frac{l}{\ell},
\label{dthdph}%
\end{equation}
where $\sigma$ is an arbitrary constant. Solving Eq. (\ref{dthdph}) gives the
duality transformations (\ref{rrho}) and (\ref{thetaphi}).

\textit{Three-dimensional power potentials.} Consider three-dimensional power
potentials as an example. The duality of a power potential, generally
speaking, is no longer a power potential. However, if requiring that the dual
potential is still a power potential, we have the following result. The power
potentials $U\left(  r\right)  =\xi r^{a}$ and $V\left(  \rho\right)
=\eta\rho^{A}$ are dual to each other, if $\frac{a+2}{2}=\frac{2}{A+2}$. The
orbit of the potential $U\left(  r\right)  $ with the energy $E$ and the orbit
of its dual potential $V\left(  \rho\right)  $ with the energy $\mathcal{E}%
$\ can be obtained from each other by the replacement, $r\leftrightarrow
\rho^{\frac{l}{\ell}\frac{2}{a+2}}$ and $\theta\leftrightarrow\frac{l}{\ell
}\frac{2}{a+2}\phi$.

\textit{The Newton-Hooke duality.} An important special case of the duality is
the Newton-Hooke duality. The Newton-Hooke duality is a duality between the
Newtonian gravitational potential and the harmonic-oscillator potential, which
is revealed by Newton in his Principia \cite{chandrasekhar1995newton}. The
Newtonian gravitational potential, in fact, has an infinite number of dual
potentials corresponding to various choices of the parameter $\sigma$. The
Newton-Hooke duality, the duality between $U\left(  r\right)  =\xi/r$ and
$V\left(  \rho\right)  =\eta\rho^{2}$, corresponds to $\sigma=2$. The energy
of the Newtonian gravitational potential system becomes the coupling constant
of its dual potential.

\subsection{The duality in quantum mechanics \label{quantum}}

The equation of motion in quantum mechanics is the Schr\"{o}dinger equation.%

\begin{tcolorbox}[boxrule=0pt,
  boxsep=0pt,
  colback={lightgray1},
  enhanced jigsaw,
  borderline west={3pt}{0pt}{lightgray2},
  sharp corners,
  before skip=10pt,
  after skip=10pt,
breakable,]

\textit{Two one-dimensional stationary Schr\"{o}dinger equations with
potentials }$U\left(  x\right)  $\textit{ and }$V\left(  \xi\right)
$\textit{,}%
\begin{align}
\frac{d^{2}u\left(  x\right)  }{dx^{2}}+\left[  E-U\left(  x\right)  \right]
u\left(  x\right)   &  =0,\label{requ1d}\\
\frac{d^{2}v\left(  \xi\right)  }{d\xi^{2}}+\left[  \mathcal{E}-V\left(
\xi\right)  \right]  v\left(  \xi\right)   &  =0,\label{veq1d}%
\end{align}
\textit{where }$E$\textit{ and }$E$\textit{\ are eigenvalues, if the
potentials }$U\left(  x\right)  $\textit{ and }$V\left(  \xi\right)  $\textit{
are related by }%
\begin{equation}
\sigma\left\{  x^{2}\left[  U\left(  x\right)  -E\right]  +\frac{1}%
{4}\right\}  =\frac{1}{\sigma}\left\{  \xi^{2}\left[  V\left(  \xi\right)
-\mathcal{E}\right]  +\frac{1}{4}\right\}  \label{rUrhoVq1d}%
\end{equation}
\textit{with}%
\begin{equation}
x\leftrightarrow\xi^{\sigma},\label{rrhoq1d}%
\end{equation}
\textit{the eigenfunctions }$u\left(  x\right)  $\textit{ and }$v\left(
\xi\right)  $\textit{ are related by the duality transformation}%
\begin{equation}
u\left(  x\right)  \leftrightarrow\xi^{\left(  \sigma-1\right)  /2}v\left(
\xi\right)  .\label{uvq1d}%
\end{equation}
\textit{Here }$\sigma$\textit{ is a constant chosen arbitrarily.}%
\end{tcolorbox}

\textit{The duality of the Poschl-Teller potential.} For the Poschl-Teller
potential $U\left(  x\right)  =\alpha\operatorname*{sech}\nolimits^{2}x$, the
stationary Schr\"{o}dinger equation has the solution, $u\left(  x\right)
=P_{\left(  \sqrt{1-4\alpha}-1\right)  /2}^{i\sqrt{E}}\left(  \tanh x\right)
$ with $P_{l}^{m}\left(  z\right)  $ the associated Legendre polynomial. The
dual potential of the\textit{ }Poschl-Teller potential is $V\left(
\xi\right)  =\frac{1}{4}\left(  \sigma^{2}-1\right)  \frac{1}{\xi^{2}}%
+\sigma^{2}\xi^{2\left(  \sigma-1\right)  }\left(  \alpha\operatorname*{sech}%
\nolimits^{2}\xi^{\sigma}-E\right)  +\mathcal{E}$ and its solution \\$v\left(
\xi\right)  =\xi^{\frac{1-\sigma}{2}}P_{\left(  \sqrt{1-4\alpha}-1\right)
/2}^{i\sqrt{E}}\left(  \tanh\xi^{\sigma}\right)  $. Different choices of
$\sigma$ give different dual potentials. The constant $\mathcal{E}$ in the
dual potential $V\left(  \xi\right)  $ can also be chosen arbitrarily, since
it is a constant added in the potential.%

\begin{tcolorbox}[boxrule=0pt,
  boxsep=0pt,
  colback={lightgray1},
  enhanced jigsaw,
  borderline west={3pt}{0pt}{lightgray2},
  sharp corners,
  before skip=10pt,
  after skip=10pt,
breakable,]
\textit{For the radial equation of }$n$\textit{-dimensional central potential
}$U\left(  r\right)  $\textit{ and the radial equation of }$m$%
\textit{-dimensional central potential }$V\left(  \rho\right)  $\textit{,}%
\begin{align}
\frac{d^{2}u_{l}\left(  r\right)  }{dr^{2}}+\left[  E-\frac{\left(  l-\frac
{3}{2}+\frac{n}{2}\right)  \left(  l-\frac{1}{2}+\frac{n}{2}\right)  }{r^{2}%
}-U\left(  r\right)  \right]  u_{l}\left(  r\right)   &  =0,\label{requ}\\
\frac{d^{2}v_{\ell}\left(  \rho\right)  }{d\rho^{2}}+\left[  \mathcal{E}%
-\frac{\left(  \ell-\frac{3}{2}+\frac{m}{2}\right)  \left(  \ell-\frac{1}%
{2}+\frac{m}{2}\right)  }{\rho^{2}}-V\left(  \rho\right)  \right]  v_{\ell
}\left(  \rho\right)   &  =0,
\end{align}
\textit{where }$E$\textit{ and }$E$\textit{\ are eigenvalues, if the
potentials }$U\left(  r\right)  $\textit{ and }$V\left(  \rho\right)
$\textit{ are related by}%
\begin{equation}
\frac{r^{2}}{\left(  l+\frac{n}{2}-1\right)  ^{2}}\left[  U\left(  r\right)
-E\right]  =\frac{\rho^{2}}{\left(  \ell+\frac{m}{2}-1\right)  ^{2}}\left[
V\left(  \rho\right)  -\mathcal{E}\right]  \label{rUrhoVq}%
\end{equation}
\textit{with}%
\begin{equation}
r\leftrightarrow\rho^{\sigma},\label{rrhoq}%
\end{equation}
\textit{the eigenfunctions }$u_{l}\left(  r\right)  $\textit{ and }$v_{\ell
}\left(  \rho\right)  $\textit{ are related by the duality transformation}%
\begin{equation}
u_{l}\left(  r\right)  \leftrightarrow\rho^{\left(  \sigma-1\right)
/2}v_{\ell}\left(  \rho\right)  .\label{uvq}%
\end{equation}
\textit{The relation between the angular momenta of the dual systems, then, is
}%
\begin{equation}
l+\frac{n}{2}-1\leftrightarrow\frac{1}{\sigma}\left(  \ell+\frac{m}%
{2}-1\right)  .\label{ltrans}%
\end{equation}
\textit{Here }$\sigma$\textit{ is a constant chosen arbitrarily.}%
\end{tcolorbox}

It should be emphasized that the duality relation of the angular momentum, Eq.
(\ref{ltrans}), is a result of the dual transformations (\ref{rrhoq}) and
(\ref{uvq}).

\textit{Three-dimensional power potentials.} A three-dimensional power
potential has infinite number of dual potentials. If requiring the dual
potential of a power potential is also a power potential, we have the
following result.

The power potentials $U\left(  r\right)  =\xi r^{a}$ and $V\left(
\rho\right)  =\eta\rho^{A}$ are dual to each other, if $\frac{a+2}{2}=\frac
{2}{A+2}$. The solution are related by the transformation $r\leftrightarrow
\rho^{2/\left(  a+2\right)  }$ and $u_{l}\left(  r\right)  \leftrightarrow
\rho^{-a/\left[  2\left(  a+2\right)  \right]  }v_{\ell}\left(  \rho\right)  $.

\textit{The Newton-Hooke duality.} In quantum mechanics there still exists the
Newton-Hooke duality, i.e., the duality between the Newtonian gravitational
potential and the harmonic-oscillator potential. The solution of radial
equation of the Newtonian gravitational potential $U\left(  r\right)  =\xi
/r$,
\begin{equation}
u_{l}\left(  r\right)  =Ae^{-\sqrt{-E}r}\left(  2\sqrt{-E}\right)
^{l+1}r^{l+1}{}_{1}F_{1}\left(  l+1+\frac{\xi}{2\sqrt{-E}},2\left(
l+1\right)  ,2\sqrt{-E}r\right)  ,
\end{equation}
and the solution of radial equation of the harmonic-oscillator potential
$V\left(  \rho\right)  =-4E\rho^{2}$,%
\begin{equation}
v_{\ell}\left(  \rho\right)  =Ae^{-\sqrt{-E}\rho^{2}}\left(  2\sqrt
{-E}\right)  ^{\frac{\ell}{2}+\frac{3}{4}}\rho^{\ell+1}{}_{1}F_{1}\left(
\frac{\ell}{2}+\frac{3}{4}+\frac{\xi}{2\sqrt{-E}},\ell+\frac{3}{2},2\sqrt
{-E}\rho^{2}\right)
\end{equation}
are related by the transformation $r\leftrightarrow\rho^{2}$, $u_{l}\left(
r\right)  \leftrightarrow\rho^{1/2}v_{\ell}\left(  \rho\right)  $, and
$l+\frac{1}{2}\leftrightarrow\frac{1}{2}\left(  \ell+\frac{1}{2}\right)  $.

For more examples see Ref. \cite{li2017quantum}.

\section{The traveling wave solution of scalar field equation
\label{travelingwavesolution}}

In this appendix we derive the solution of the field equation, Eq.
(\ref{solphia}), in section \ref{Solving}.

\textit{Approach 1.} The traveling wave solution is a solution satisfying%
\begin{equation}
\phi\left(  x^{\mu}\right)  =\phi\left(  \beta_{\mu}x^{\mu}\right)
=\phi\left(  z\right)  ,
\end{equation}
where $z=\beta_{\mu}x^{\mu}$. A familiar special case is the $1+1$-dimensional
case with $\beta_{\mu}=\left(  1,-1\right)  $ and in this case $z=t-x$.

Substituting%
\begin{equation}
\square\phi=\partial^{\mu}\partial_{\mu}\phi=\frac{\partial z}{\partial
x_{\mu}}\frac{d}{dz}\left(  \frac{\partial z}{\partial x^{\mu}}\frac{d\phi
}{dz}\right)  =\beta^{2}\frac{d^{2}\phi}{dz^{2}}%
\end{equation}
into the field equation, $\square\phi+m^{2}\phi+\frac{\partial V\left(
\phi\right)  }{\partial\phi}=0$, gives%
\begin{equation}
\beta^{2}\frac{d^{2}\phi}{dz^{2}}+m^{2}\phi+\frac{\partial V\left(
\phi\right)  }{\partial\phi}=0.
\end{equation}

Multiplying both sides by $\frac{d\phi}{dz}$ and integrating over $z$ give
\begin{equation}
\frac{1}{2}\beta^{2}\left(  \frac{d\phi}{dz}\right)  ^{2}+\frac{1}{2}m^{2}%
\phi^{2}+V\left(  \phi\right)  -c=0, \label{solution1}%
\end{equation}
where $c$ is the integration constant. By Eq. (\ref{solution1}) we have%
\begin{equation}
dz=d\phi\frac{\sqrt{-\beta^{2}}}{\sqrt{2\left[  \frac{1}{2}m^{2}\phi
^{2}+V\left(  \phi\right)  -c\right]  }}.
\end{equation}
Integrating both sides gives%
\begin{equation}
\beta_{\mu}x^{\mu}+\int\frac{\sqrt{-\beta^{2}}}{\sqrt{2\left[  \frac{1}%
{2}m^{2}\phi^{2}+V\left(  \phi\right)  -c\right]  }}d\phi=0; \label{xingbo}%
\end{equation}
note that $\beta_{\mu}x^{\mu}=z$.

Rewriting Eq. (\ref{xingbo}) as%
\begin{equation}
\beta_{\mu}x^{\mu}=-\int_{\phi_{0}}^{\phi}\frac{\sqrt{-\beta^{2}}}%
{\sqrt{2\left[  \frac{1}{2}m^{2}\varphi^{2}+V\left(  \varphi\right)
-c\right]  }}d\varphi
\end{equation}
and taking the derivative with respect to $x^{\mu}$ give%
\begin{equation}
\partial_{\mu}\phi=-\beta_{\mu}\frac{\sqrt{2\left[  \frac{1}{2}m^{2}\phi
^{2}+V\left(  \phi\right)  -c\right]  }}{\sqrt{-\beta^{2}}}. \label{dphi}%
\end{equation}
Substituting Eq. (\ref{dphi}) into the expression of $G$, Eq. (\ref{Ephi}),
gives%
\begin{equation}
G=c.
\end{equation}
Substituting $c=G$ into the solution (\ref{xingbo}) gives the solution
(\ref{solphia}), i.e.,
\begin{equation}
\beta_{\mu}x^{\mu}+\int\frac{\sqrt{-\beta^{2}}}{\sqrt{2\left(  \frac{1}%
{2}m^{2}\phi^{2}+V\left(  \phi\right)  -G\right)  }}d\phi=0. \label{Tsolution}%
\end{equation}

\textit{Approach 2. }Alternatively, we can also directly verify that\ Eq.
(\ref{Tsolution}) is a solution of the field equation.

Rewrite Eq. (\ref{Tsolution}) as%
\begin{equation}
F_{1}\left(  x^{\mu},\phi\right)  =0, \label{gensol}%
\end{equation}
where
\begin{equation}
F_{1}\left(  x^{\mu},\phi\right)  \equiv\beta_{\mu}x^{\mu}+\int\frac
{\sqrt{-\beta^{2}}}{\sqrt{2\left[  \frac{1}{2}m^{2}\phi^{2}+V\left(
\phi\right)  -G\right]  }}d\phi.
\end{equation}
By the formula for derivative of implicit function, we have%
\begin{equation}
\partial_{\mu}\phi=-\displaystyle\frac{\frac{\partial F_{1}\left(  x^{\mu
},\phi\right)  }{\partial x^{\mu}}}{\frac{\partial F_{1}\left(  x^{\mu}%
,\phi\right)  }{\partial\phi}}=-\frac{\beta_{\mu}\sqrt{2\left[  \frac{1}%
{2}m^{2}\phi^{2}+V\left(  \phi\right)  -G\right]  }}{\sqrt{-\beta^{2}}}.
\label{rdphi1}%
\end{equation}

Rewrite Eq. (\ref{rdphi1}) as
\begin{equation}
F_{2}\left(  \phi,\partial_{\mu}\phi\right)  =0,
\end{equation}
where
\begin{equation}
F_{2}\left(  \phi,\partial_{\mu}\phi\right)  =\partial_{\mu}\phi+\frac
{\beta_{\mu}\sqrt{2\left[  \frac{1}{2}m^{2}\phi^{2}+V\left(  \phi\right)
-G\right]  }}{\sqrt{-\beta^{2}}}.
\end{equation}
By the formula for derivative of implicit function, we have
\begin{align}
\square\phi &  =\partial^{\mu}\left(  \partial_{\mu}\phi\right)  =\left[
\frac{\partial}{\partial\phi}\left(  \partial_{\mu}\phi\right)  \right]
\frac{\partial\phi}{\partial x_{\mu}}=-\frac{\frac{\partial F_{2}\left(
\phi,\partial_{\mu}\phi\right)  }{\partial\phi}}{\frac{\partial F_{2}\left(
\phi,\partial_{\mu}\phi\right)  }{\partial\left(  \partial_{\mu}\phi\right)
}}\partial^{\mu}\phi\nonumber\\
&  =-\left(  m^{2}\phi+\frac{\partial V\left(  \phi\right)  }{\partial\phi
}\right)  ,
\end{align}
where $\frac{\partial F_{2}\left(  \phi,\partial_{\mu}\phi\right)  }%
{\partial\phi}=-\left(  m^{2}\phi+\frac{\partial V\left(  \phi\right)
}{\partial\phi}\right)  $, $\frac{\partial F_{2}\left(  \phi,\partial_{\mu
}\phi\right)  }{\partial\left(  \partial_{\mu}\phi\right)  }=1$, and note that
$\partial^{\mu}\phi=\frac{\partial\phi}{\partial x_{\mu}}$.

This proves that Eq. (\ref{Tsolution}) is a solution of the field equation.

\bigskip
\bigskip
\bigskip



\acknowledgments

We are very indebted to Dr G. Zeitrauman for his encouragement. This work is supported in part by Special Funds for theoretical physics Research Program of the NSFC under Grant No.
11947124, and NSFC under Grant Nos. 11575125 and 11675119.

\nolinenumbers

\reftitle{References}




\providecommand{\href}[2]{#2}\begingroup\raggedright\endgroup

\end{paracol}
%


\end{document}